\documentclass{aa}

\input psfig.sty
\input epsf.sty

\newcommand{\gapprox}{\raisebox{-0.4ex}{$\gppr $}}
\newcommand{\gppr}{\stackrel{>}{\scriptstyle \sim}}

\begin{document}

\title{A large Wolf-Rayet population in NGC~300 uncovered by VLT-FORS2
\thanks{Based on observations made with ESO Telescopes at the Paranal 
Observatory under programme ID 65.H-0705(A)}} 

\titlerunning{WR stars in NGC~300}

% \subtitle{}

\author{H.~Schild\inst{1} \and
P.A.~Crowther\inst{2} \and 
J.B.~Abbott\inst{2} \and
W.~Schmutz\inst{3}
}

\authorrunning{Schild et al.}

     \institute{Institut f\"ur Astronomie, ETH-Zentrum, CH 8092 Z\"urich, 
           	Switzerland
\and Dept of Physics and Astronomy, University College London, 
Gower St, London WC1E 6BT, United Kingdom
\and        
     Physikalisch-Meteorologisches Observatorium, CH-7260 Davos,
           Switzerland
}
\offprints{P.A. Crowther, \email{pac@star.ucl.ac.uk}}
\date{Received / Accepted}

\abstract{We have detected 58 Wolf-Rayet candidates in
the central region of the nearby spiral galaxy
NGC~300, based on deep VLT-FORS2 narrow-band imaging.
Our survey is close to complete except for heavily reddened
WR stars.
Of the objects in our list, 16 stars were already spectroscopically 
confirmed as WR stars by Schild \& Testor and Breysacher et al., 
to which 4 stars are added using low resolution FORS2 datasets. 
The WR population of NGC~300 now totals 60, a threefold 
increase over previous surveys, with WC/WN$\geq$1/3, in reasonable
agreement with Local Group galaxies for a moderately sub-solar metallicity.
We also discuss the WR surface density in the central 
region of NGC~300. Finally, analyses are presented for 
two apparently single WC stars -- \#29 (alias WR3, WC5) and \#48 (alias
WR13, WC4) located close to the nucleus, and at a deprojected radius of
2.5~kpc, respectively. These are among the first models of WR stars in 
galaxies beyond the Local  Group, and are compared with early WC stars in 
our Galaxy and LMC.
}

\maketitle
%\markboth{}{}

\section{Introduction}

Over 500 Wolf-Rayet stars have been identified in Local Group
galaxies, principally the Milky Way, M31 and M33. These stars beautifully
trace young stellar populations, and their number and distribution reacts 
sensitively to metallicity, which varies by an order of magnitude from the 
Small Magellanic Cloud (SMC) to M31. Detailed studies of individual WR stars
in Local Group stars have been carried out (e.g. Smartt et al. 2001; 
Crowther 2000; Crowther et al. 2002) using 2--4m class telescopes.

The availability of 8--10m class telescopes
permits the discovery and study of individual stars at greater distances,
spanning a greater range of metallicities. As a first application, we
present here VLT imaging and spectroscopy of WR stars in NGC~300, located
in the Sculptor group at a distance of 2 Mpc (Freedman et al. 2001). It's
metallicity is bracketed by the Milky Way and Large 
Magellanic Cloud (LMC) and therefore we expect a similarly large number
of WR stars in NGC~300. Previous surveys have however failed to identify them. 
A large population might also be
anticipated since NGC~300 is a late type spiral, reminiscent of M33, 
which harbours at least 140 WR stars (Massey \& Johnson 1998).
Because of its low inclination NGC~300 is well suited to studies of its 
stellar 
content -- recent surveys include blue supergiants 
(Bresolin et al. 2002a), Cepheids (Pietrzy\'nski
et al. 2002b), OB associations (Pietrzy\'nski et al. 2001) and 
Supernova 
remnants (Pannuti et al. 2001). 

The first signature of WR stars in NGC\,300 was 
found in spectra of H\,{\sc ii} regions by D'Odorico et al. 
(1983). They detected the broad WR feature in two out of sixteen 
H\,{\sc ii} 
regions. Five years later, Deharveng et al. (1988) presented a catalogue of 
176 H\,{\sc ii} regions and found broad WR emission in four of them. 
Although this clearly demonstrated that WR detection was feasible at
this distance, the spectroscopy could not in itself locate the individual
WR stars. To achieve this, imaging in narrow band filters was necessary.
First results with this technique were reported by Schild \& Testor (1991,
1992) and Testor \& Schild (1993). They found in total 13 WR candidates and 
confirmed them spectroscopically. Six additional WR stars were later
identified in the same way by Breysacher et al. (1997) in stellar 
associations. One additional weak--lined late WN star was 
serendipitously found by Bresolin et al. (2002ab).

\begin{figure*}
\mbox{\psfig{figure=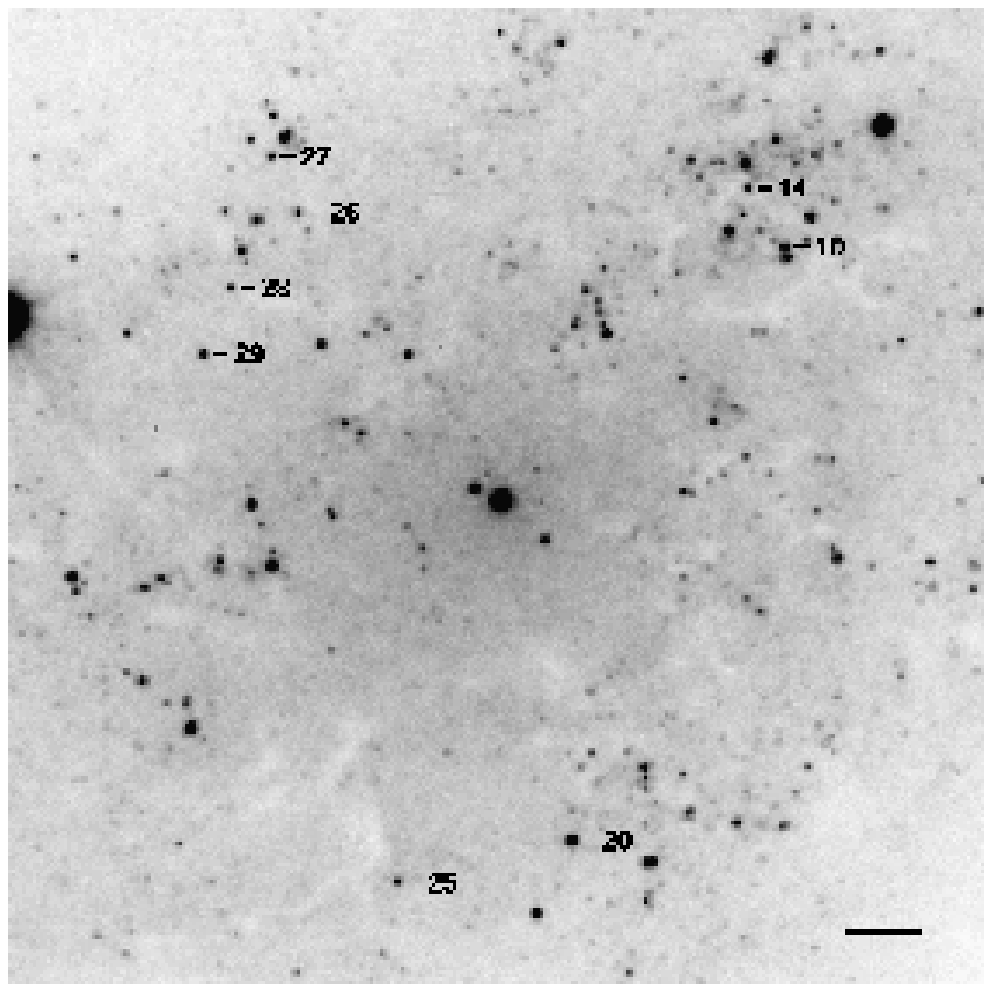,width=8.5cm,height=8.5cm,clip=}}
\hskip 0.8cm
\mbox{\psfig{figure=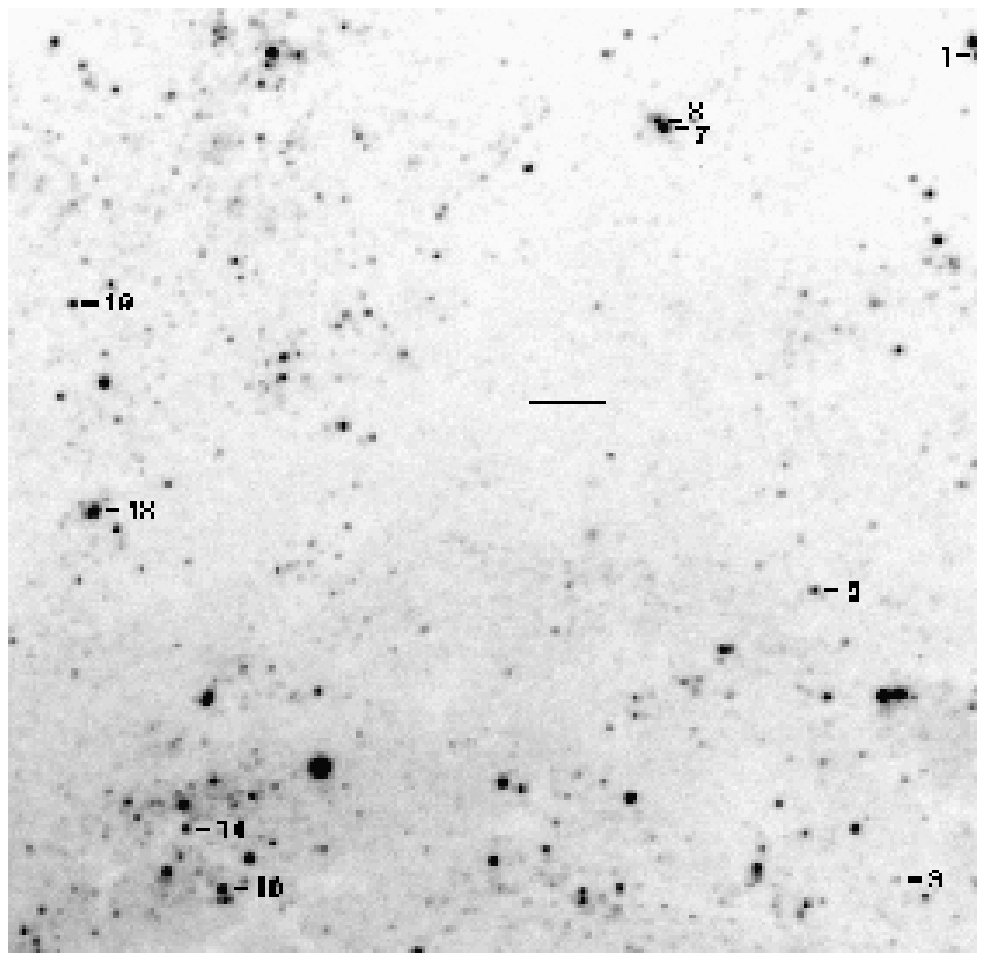,width=8.5cm,height=8.5cm,clip=}}

\caption[]{Finding chart for WR stars/candidates in
NGC~300. Left: nuclear region, right: area northwest of the nucleus. 
The horizontal bar represents
10\,\arcsec. North to the top, East to the left.}
\label{nucnw.fig}
\end{figure*}

In total, there are presently 22 confirmed WR stars in 
NGC~300, strongly skewed towards WC subtypes. 
From a census of the WR distribution  in 
Local Group galaxies, Massey (1996) identified a rather tight
correlation between the WC/WN ratio and metallicity, as characterized 
by oxygen content. 
Although there have not been any recent studies of the
NGC~300 metallicity gradient, Deharveng et al. (1988) used data from
Pagel et al. (1979) and  Webster \& Smith (1983), to imply
a range between log(O/H)+12=8.9 in its nucleus and  8.3 
in its outer spiral arms. Similar conclusions were obtained by 
Zaritsky et al. (1994) from a recalibration of previous results.
 One would expect a WC/WN ratio of $\sim$\,1/2 
from comparison with Local 
Group galaxies, yet the census of WR stars in NGC~300 indicates 
WC/WN\,$\sim$\,2. Consequently, we might expect 
that the WR population of NGC~300 is highly incomplete, particularly
amongst WN stars.

In this paper we  present results from a new imaging survey of the
central region of NGC\,300 with the Very Large Telescope (VLT). New WR 
candidates are identified, some of which are spectroscopically confirmed. 
Spectral types of the latter are discussed, with particular reference to 
the WC/WN ratio of the inner galaxy. An analysis of two apparently 
single WC stars is presented, one located close to its nucleus,
the other at $\sim$\,50\,\% of the Holmberg radius, $\rho_0$. 
Comparisons are made with 
recent comparable studies of WC stars in a variety of metallicity 
environments. 

\section{Observations}

We observed NGC\,300 with the VLT UT2 (Kueyen) and Focal Reduced/Low
Dispersion Spectrograph \#2 (FORS2) during 2000 September 2--3. 
The conditions were photometric but the seeing was highly variable,
changing from 0.6 to 3.5\,\arcsec\ and 
we used the instrument accordingly in imaging and spectroscopic mode.

\subsection{Imaging}

While the seeing was good (typically 0.8\,\arcsec)
we obtained images through two interference filters
with central wavelengths at 4684\,\AA\ and 4781\,\AA\ and band widths of
66\,\AA\ and 68\,\AA, respectively. The former filter is well matched to
the strong WR emission feature containing the
N\,{\sc iii}
$\lambda$4640, C\,{\sc iii} $\lambda$4650, C\,{\sc iv} 
$\lambda$4658 and  He\,{\sc ii} $\lambda$4686 emission lines.
The wavelength range of the latter filter falls into a
spectral region that is free from emission lines. We collected two
images in each filter with exposure times of 600\,sec. These frames
were centred at $\alpha$: 0\,h 54\,m 59.0\,s and $\delta$: --37\,$^{\circ}$ 
40\,\arcmin\ 59\,\arcsec\ (2000).
At mediocre seeing conditions short exposures through Bessel B and V filters
were also collected. Only one of the V frames was of sufficient quality, 
but it was slightly offset such that only V-band magnitudes 
of WR candidates
with RA larger than 00\,h 54\,m 46.7\,s could be measured.

The standard collimator was used, providing a field-of-view of
6.8\,\arcmin$\times$6.8\,\arcmin\ with an image scale of 0.2\,\arcsec/pixel. 
The detector was a 2048$\times$2048 Tektronix CCD with 24\,$\mu$m pixels.

The data were de-biased and flat fielded with frames taken in the 
following morning twilight. We used the DAOPHOT software package to get
relative photometry. These were converted into absolute fluxes with
the photometric standard stars in the field of NGC 300 listed in
Pietrzi\'nski et al. (2002a).

\begin{table*}
\caption[]{List of Wolf-Rayet stars and candidates in the central
regions of NGC~300. Two other WR stars are known outside the present
field -- 
WR8 (WN) in Deh~30 from Schild \& Testor (1992) and a WC in Deh~24 
from D'Odorico et al. (1983).
Deprojected galactocentric distances are expressed 
as a fraction of the Holmberg radius ($\rho_0$=9.75$' \simeq$ 5.75 kpc).}
\label{catalog}
\begin{tabular}{rllrcrrrc|lll}
\noalign{\hrule\smallskip}
   &     RA (J2000)     &     Dec (J2000)     
&$m_{4684}$&$m_{4781}-m_{4684}$&$m_{V}$ &
$\rho/\rho_{0}$ & Deh$^{4)}$ 
& Spect Type & ST$^{1)}$ & Brey$^{2)}$ & 
Remark\cr
\noalign{\smallskip\hrule\smallskip}
%   a &  0 54 28.7~ & -37 41 33~  &        &        &     & 0.52 && WC    & \cr %d83
%   b &  0 54 31.3~ & -37 37 59~  &        &        &     & 0.68 && WNE   &WR8 st92 &   \cr
   1 &  0 54 41.7~ & -37 38 39.0  & out    &  out   &out  & 0.44 &48& WC4   &&& $^{5)}$\cr
   2 &  0 54 42.37 & -37 43  7.3 & 18.93  &  1.02  & out & 0.33 &53b& WC4   &WR\,10 &    \cr  
   3 &  0 54 42.61 & -37 40 28.5 & 21.88  &  1.25  & out & 0.25 &&       &       &\cr % \#9*  
   4 &  0 54 42.78 & -37 43  1.8 & 17.88  &  0.25  & out & 0.32 &53c& WN    &       && $^{4)}$ \cr % \#13* 
   5 &  0 54 43.53 & -37 39 49.9 & 20.52  &  2.09  & out & 0.28 &54&       &       &\cr % \#8 
   6 &  0 54 44.75 & -37 42 40.0 & 19.03  &  0.13  & out & 0.26 && WN11  & &&
$^{3)}$ \cr
   7 &  0 54 45.22 & -37 38 48.0 & 19.71  &  0.14  & out & 0.37 &61\cr
   8 &  0 54 45.28 & -37 38 47.0 & 20.22  &  0.24  & out & 0.37 &61\cr
   9 &  0 54 47.69 & -37 42 45.3 & 20.26  &  0.58  & 20.62 & 0.24 && WN4--5  &&& 
$^{5)}$ \cr % \#2
  10 &  0 54 50.21 & -37 40 29.7 & 19.68  &  0.52  & 20.04 & 0.10 &76a&     & \cr % \#10*
  11 &  0 54 50.22 & -37 38 24.1 & 18.51  &  0.44  & 18.79 & 0.37 &77&WC+WN&WR\,9  & I-1 \cr
  12 &  0 54 50.53 & -37 38 26.7 & 20.87  &  1.57  & 22.12 & 0.36 &77& &       & I-2 \cr
  13 &  0 54 50.57 & -37 38 13.2 & 20.54  &  0.29  & 20.58 & 0.39 &&     &       & \cr %\#12*
  14 &  0 54 50.62 & -37 40 21.7 & 20.39  &  1.49  & 21.64 & 0.11 &76b&WC4-6&WR\,1 \cr
  15 &  0 54 50.91 & -37 38 30.0 & 20.35  &  0.16  & 20.41 & 0.35 &79\cr
  16 &  0 54 51.00 & -37 38 26.1 & 20.10  &  0.10  & 20.27 & 0.36 &79\cr
  17 &  0 54 51.31 & -37 38 26.3 & 22.50  &  0.14  & weak  & 0.36 &79\cr
  18 &  0 54 51.65 & -37 39 39.2 & 19.11  &  0.10  & 18.79 & 0.19 &84\cr
  19 &  0 54 51.89 & -37 39 11.5 & 20.20  &  0.08  & 19.24 & 0.25 &\cr
  20 &  0 54 52.59 & -37 41 49.1 & 18.54  &  0.11  & 17.57 & 0.11 &85\cr
  21 &  0 54 52.93 & -37 38 38.3 & 20.85  &  0.12  & 20.97 & 0.32 &\cr
  22 &  0 54 53.09 & -37 43 34.0 & 20.98  &  1.16  & 21.71 & 0.35 &87& WC4 &&& $^{5)}$ \cr%\#3
  23 &  0 54 53.11 & -37 43 47.3 & 19.72  &  0.08  & 19.71 & 0.38 &88\cr
  24 &  0 54 53.80 & -37 43 47.2 & 20.00  &  1.55  & 21.34 & 0.38 &90& WC5 & WR\,11 \cr
  25 &  0 54 54.57 & -37 41 54.7 & 20.28  &  0.09  & 20.27 & 0.13 &\cr
  26 &  0 54 55.67 & -37 40 25.1 & 20.12  &  0.08  & 20.10 & 0.09 &\cr
  27 &  0 54 55.99 & -37 40 17.6 & 20.39  &  0.15  & 20.47 & 0.11 &96\cr
  28 &  0 54 56.45 & -37 40 35.1 & 20.55  &  1.90  & 22.05 & 0.09 &98& WNE  & WR\,2 \cr
  29 &  0 54 56.76 & -37 40 44.0 & 19.71  &  2.58  & 21.61 & 0.08 &98& WC4-5& WR\,3 \cr
  30 &  0 54 58.95 & -37 43 58.7 & 21.24  &  1.31  & 22.01 & 0.44 &107&WN    &&& $^{5)}$ \cr%\#7
  31 &  0 55 ~0.65 & -37 38 51.5 & 20.74  &  1.68  & 22.14 & 0.31 &&      &       &\cr %\#6  
  32 &  0 55 ~2.33 & -37 38 27.4 & 18.84  &  0.08  & 18.84 & 0.37 &115\cr
  33 &  0 55 ~2.88 & -37 43 16.0 & 21.28  &  1.02  & 21.51 & 0.40 &&  &       & II-1 \cr
  34 &  0 55 ~3.34 & -37 42 42.0 & 20.36  &  1.97  & 21.87 & 0.35 &&WC5-6 &WR\,12 & III-1\cr
  35 &  0 55 ~3.55 & -37 42 49.4 & 19.06  &  0.09  & 18.82 & 0.36 &&\cr
  36 &  0 55 ~3.64 & -37 43 20.0 & 18.06  &  0.13  & 17.97 & 0.42 &\cr
  37 &  0 55 ~3.75 & -37 42 51.6 & blend  &   2.0  & 18.56 & 0.37 &\cr
  38 &  0 55 ~4.09 & -37 43 18.9 & 19.07  &  0.64  & 19.53 & 0.43 &&WN9-10&WR\,7  & II-2 \cr 
  39 &  0 55 ~4.17 & -37 43 16.6 & 20.05  &  0.08  & 19.99 & 0.42 &\cr
  40 &  0 55 ~5.69 & -37 41 13.4 & 20.13  &  1.68  & 21.56 & 0.28 && WCE  &WR\,6  & \cr
  41 &  0 55 ~9.98 & -37 42 12.5 & 21.64  &  0.95  & 22.44 & 0.43 &&      &       &  \cr %\#14* 
  42 &  0 55 11.03 & -37 37 55.5 & 22.02  &  0.56  & 22.39 & 0.52 &&      &       &  \cr %\#11* 
  43 &  0 55 12.07 & -37 41 21.9 & 19.84  &  0.12  & 19.83 & 0.42 &137d\cr
  44 &  0 55 12.19 & -37 41 19.7 & 17.71  &  0.19  & 17.80 & 0.42 &137d\cr
  45 &  0 55 12.21 & -37 41 20.4 & 18.55  &  0.40  & 18.79 & 0.42 &137d&  &       & IV-1 \cr
  46 &  0 55 12.32 & -37 41 38.4 & 20.53  &  1.89  & 21.69 & 0.44 &137a&WC4-5 &       & V-1 \cr
  47 &  0 55 12.41 & -37 41 29.0 & 19.68  &  2.95  & 21.19 & 0.43 &137b&WC5-6 &WR\,5  & IV-2 \cr
  48 &  0 55 12.54 & -37 41 23.6 & 20.23  &  3.31  & 21.87 & 0.43 &137b&WC4   &WR\,13 & IV-3 \cr
  49 &  0 55 12.58 & -37 41 39.5 & 18.46  &  0.31  & 18.85 & 0.45 &137a&WN7   &WR\,4  & V-2 \cr
  50 &  0 55 12.65 & -37 41 38.5 & 19.62  &  0.18  & 19.73 & 0.45 &137a\cr
  51 &  0 55 12.72 & -37 41 44.6 & blend  &  0.7   & 20.38 & 0.45 &137a\cr
  52 &  0 55 13.18 & -37 38  1.4 & 22.16  &  0.5   & 22.74 & 0.54 &&      &       &\cr %\#4 
  53 &  0 55 13.23 & -37 41 39.8 & 20.63  &  1.12  & 21.54 & 0.46 &&WN4-6 &       & V-3 \cr
  54 &  0 55 13.36 & -37 41 30.3 & 20.22  &  0.30  & 20.39 & 0.46 &&\cr
  55 &  0 55 13.47 & -37 41 46.3 & 20.64  &  1.87  & 22.30 & 0.47 &&WN4-5 &       & V-4 \cr
  56 &  0 55 13.51 & -37 41 37.9 & 17.73  &  0.11  & 17.68 & 0.46 &137c\cr
  57 &  0 55 13.61 & -37 41 32.0 & 20.31  &  0.11  & 20.34 & 0.46 &\cr
  58 &  0 55 13.93 & -37 41 43.2 & 18.82  &  0.08  & 18.75 & 0.48 &137c\cr
\noalign{\smallskip\hrule\smallskip}
\end{tabular}
\\
$^{1)}$: Schild \& Testor (1991, 1992), Testor \& Schild (1993);
$^{2)}$: Breysacher et al. (1997);
$^{3)}$: Bresolin et al. (2002ab); 
$^{4)}$: Deharveng et al. (1988); 
$^{5)}$: This paper
\end{table*}

\begin{figure*}
\hbox{\psfig{figure=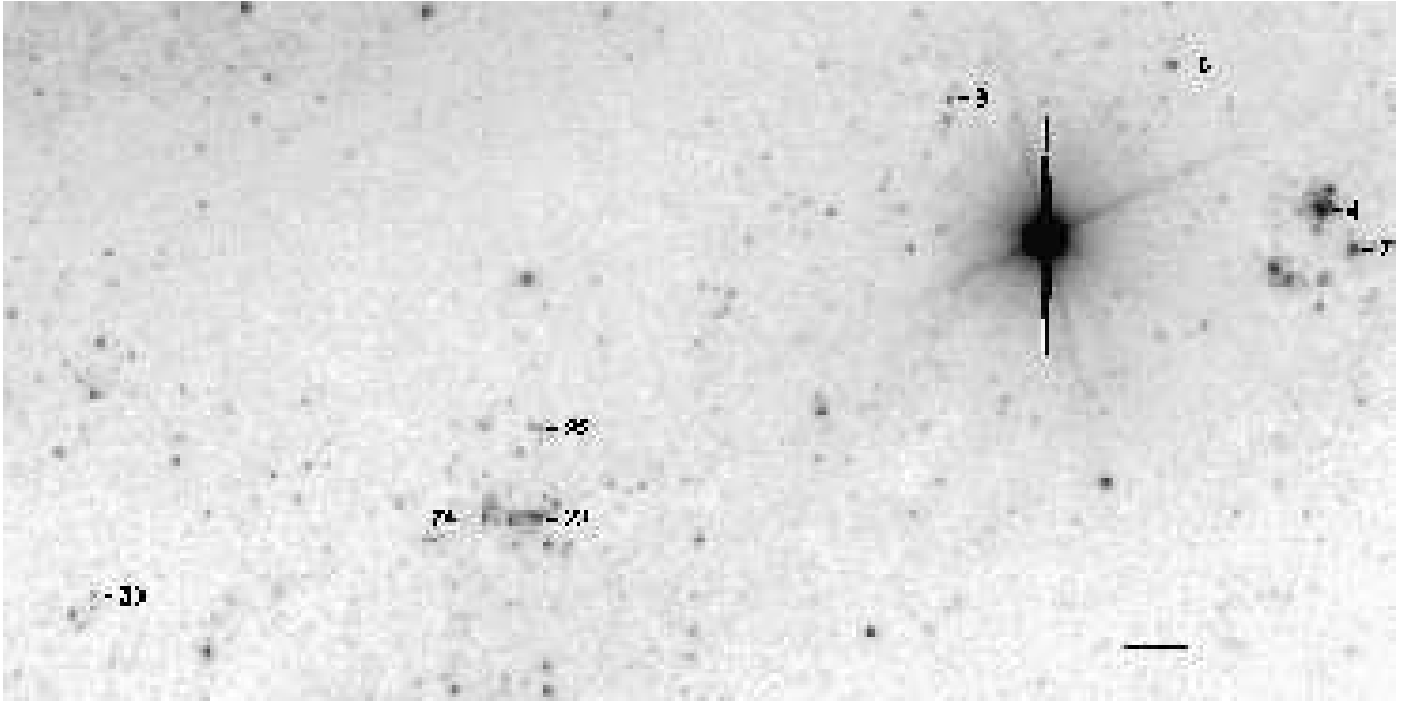,width=18.5cm,height=9.3cm,clip=}}
\hskip-0.5cm
\caption[]{Finding chart for WR stars/candidates in the southern spiral arm. The 
saturated object is the galactic foreground star CD$-$38$^{\circ}$ 301.
 The horizontal bar represents 10\,\arcsec. North to the top, East to the left.}
\label{southstrip.fig}
\end{figure*}

\begin{figure*}
\hbox{\psfig{figure=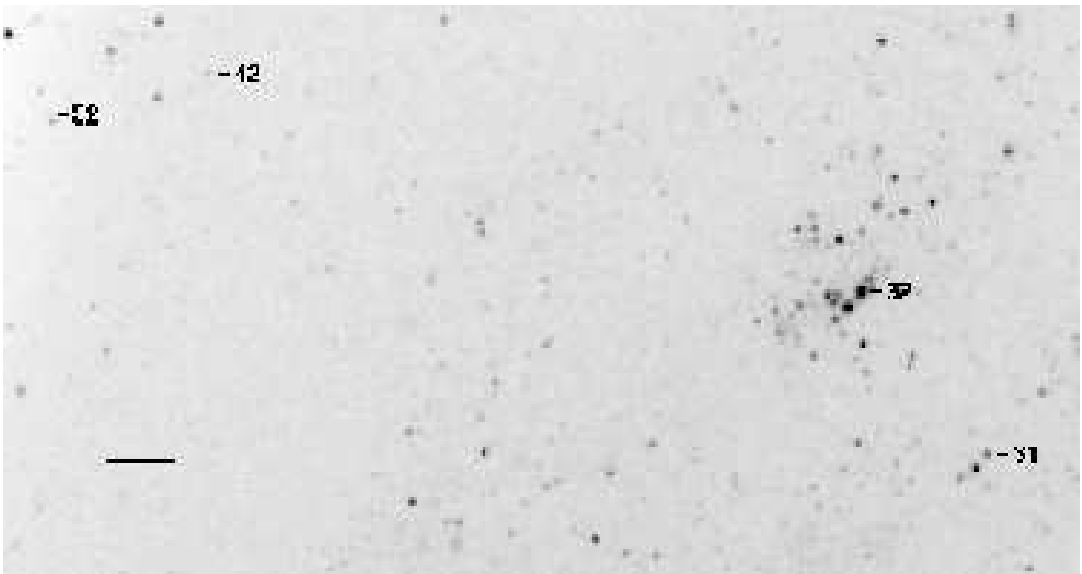,width=14.3cm,height=7.6cm,clip=}}
\hskip-0.5cm
\caption[]{Finding chart for WR stars/candidates in the northeast of 
the nucleus.
Some vignetting occurred in the upper left corner (northeast).
 The horizontal bar represents
10\,\arcsec. North to the top, East to the left.}
\label{northstrip.fig}
\end{figure*}

\begin{figure*}
\mbox{\psfig{figure=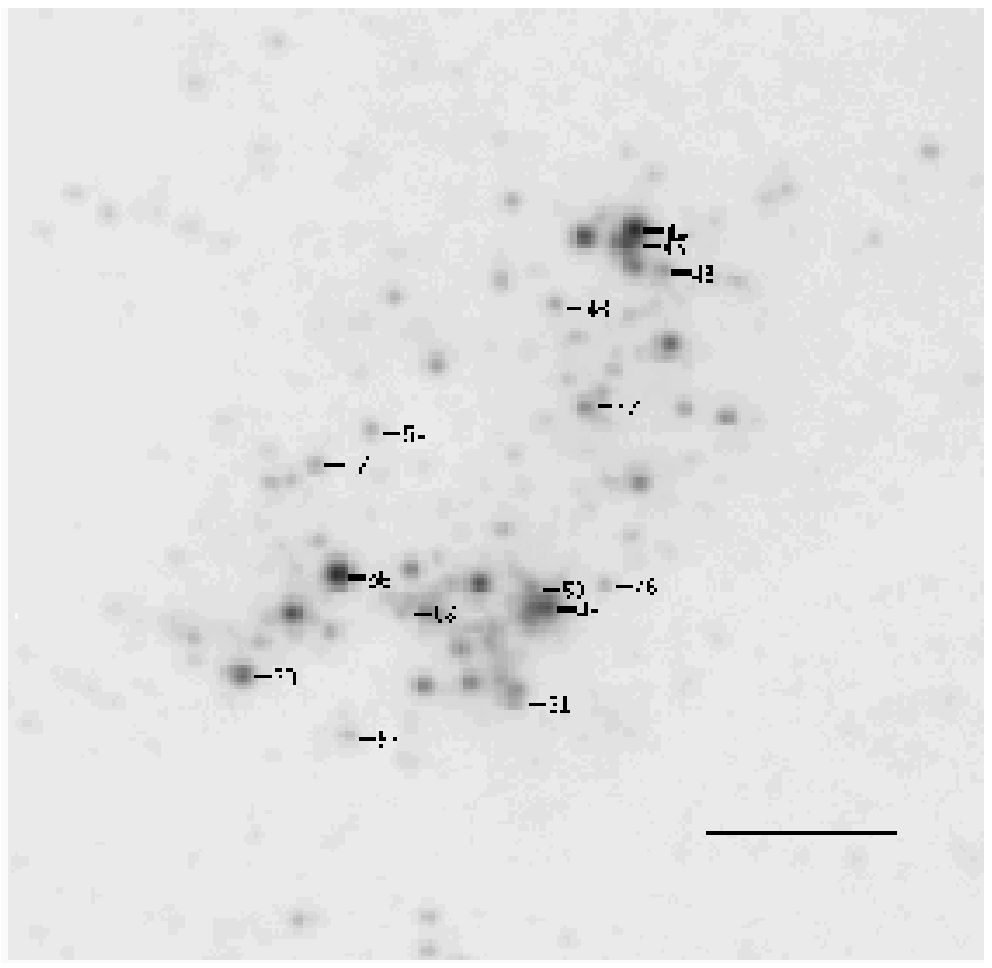,width=8.5cm,height=8.5cm,clip=}}
\vskip -8.5cm
\hskip 9.2cm
\mbox{\psfig{figure=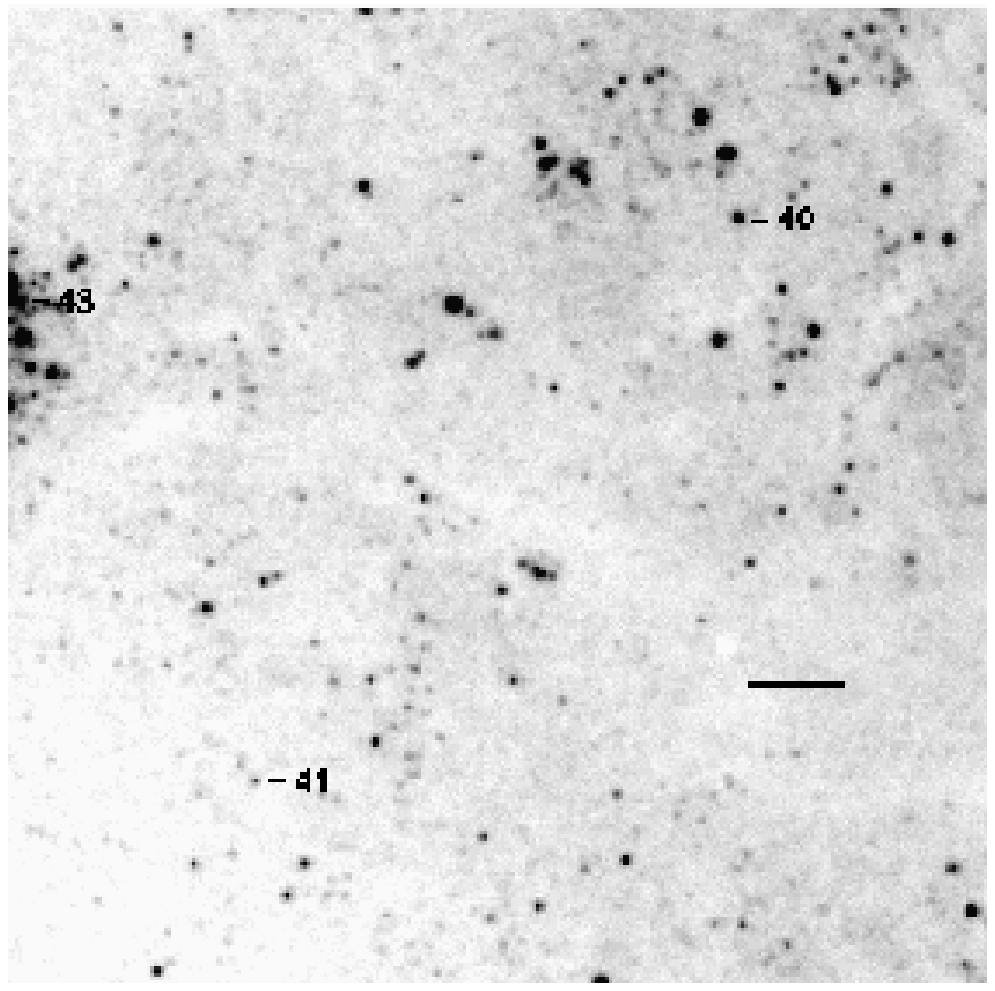,width=8.5cm,height=8.5cm,clip=}}
\vskip 1cm
\mbox{\psfig{figure=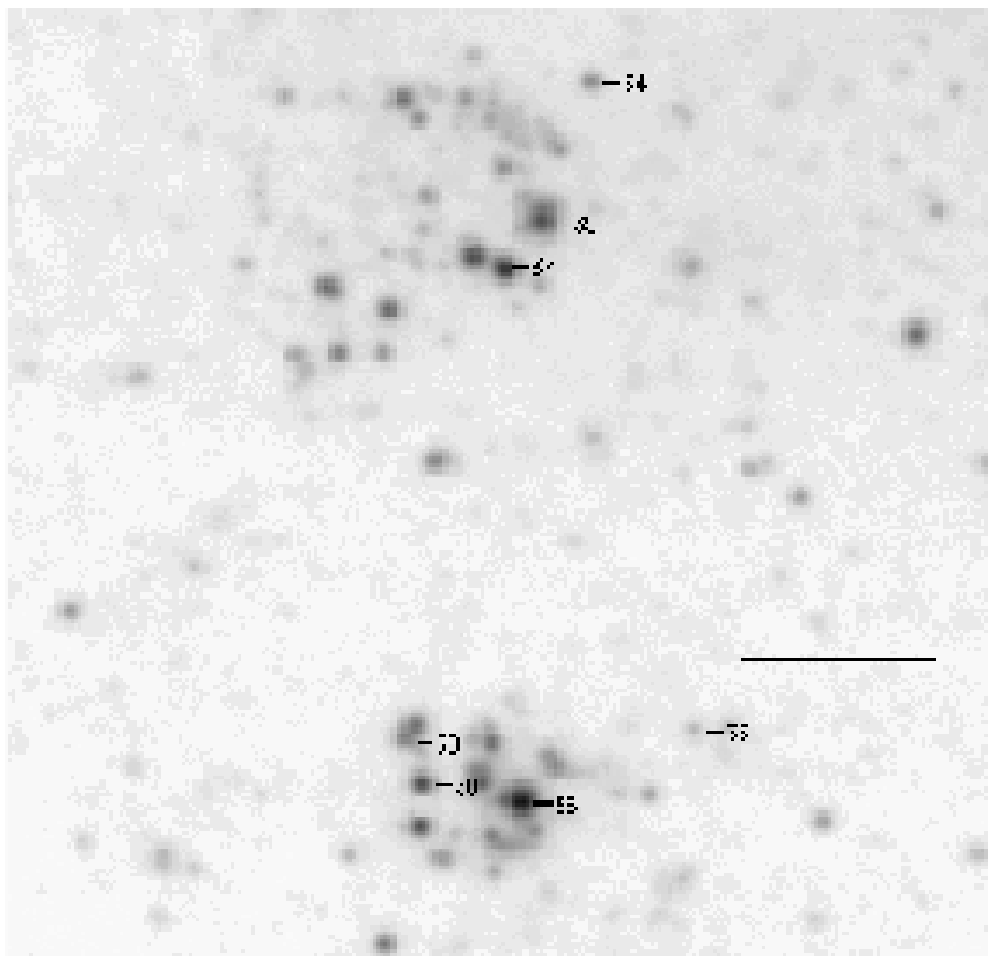,width=8.5cm,height=8.5cm,clip=}}
\vskip -8.5cm
\hskip 9.2cm
\mbox{\psfig{figure=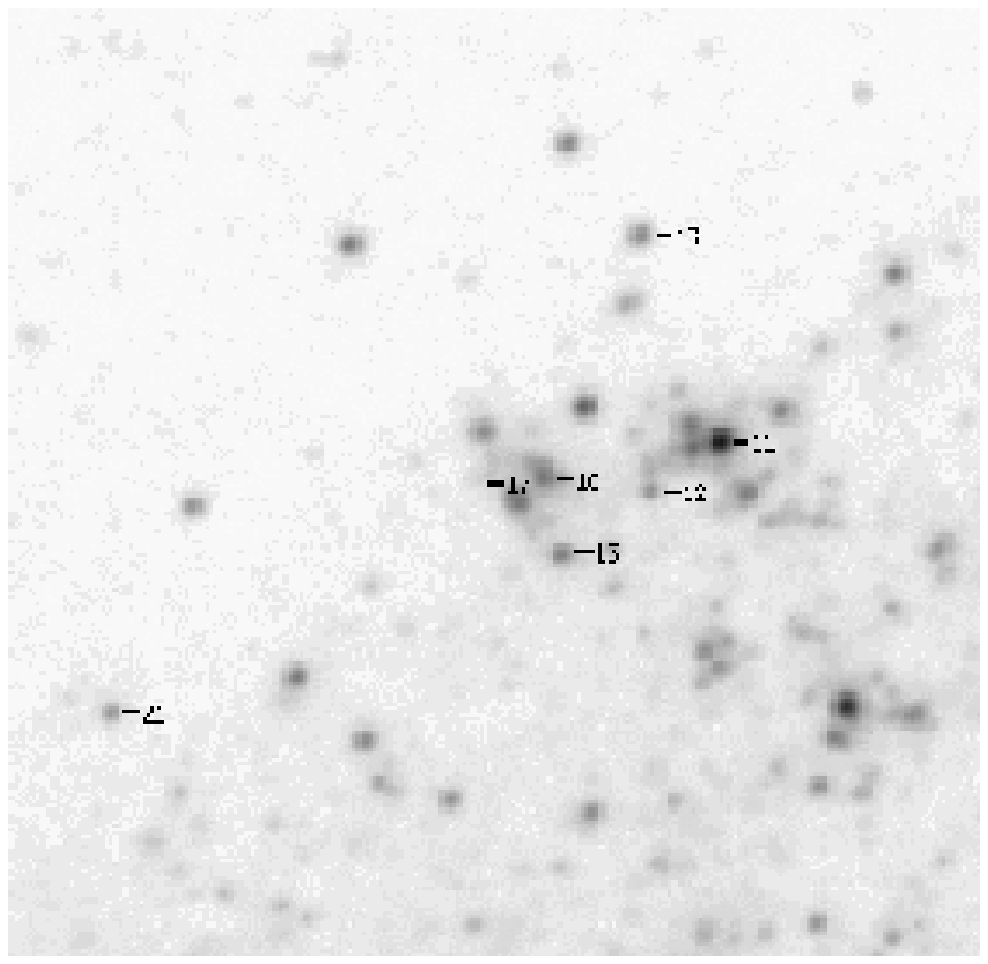,width=8.5cm,height=8.5cm,clip=}}
\hskip-0.5cm
\caption[]{Finding chart for WR stars/candidates east of the nucleus (top right)
and int he  H\,{\sc ii} regions Deh 137 (top left), Deh 118/119 (bottom left) and
Deh 77/79 (bottom right). The horizontal bars represent
10\,\arcsec. North to the top, East to the left.}
\label{hiififi.fig}
\end{figure*}

\subsection{Spectroscopy}

FORS2 was used in long slit mode (LSS) to obtain spectroscopy for
selected Wolf-Rayet stars and candidates. This mode was selected
since 
%Since the
narrow-band images were not obtained in advance of the observing run.
%we used the long slit mode (LSS) of FORS2 in preference to one of its 
%multi object mode. 
The 300V grating, GG435 filter 
and 1.0$''$ slit width provided spectroscopy covering 
$\lambda\lambda$3500--8970 at a dispersion of 1.7\,\AA/pixel, corresponding
to a resolution of R=$\Delta \lambda/\lambda\,\sim$\,440 at 5900\,\AA. 
Eight targets
% \#1, \#9, \#22, \#24, \#29, \#30, \#40 and \#48, 
were observed with this configuration using 1800 sec exposures, 
and will be discussed in 
Sect.~\ref{spectroscopy}. Generally, two WR stars were observed
in each observation via a suitable choice of position angle using
 the FORS Instrument Mask Simulator (FIMS) software.
Relative flux calibration was achieved using short exposures for 
standard stars Feige~110 and LTT~1788. Absolute calibration required
convolution with $b$ and $v$ (Smith 1968) narrow-band filters, which were
approximated to our $m_{4781}$ photometry.

For three stars,
% \#29, \#40 and \#48, 
%corresponding to Schild \& Testor WR6, 3 and 13,
 higher resolution 
600R grating observations were obtained, using the GG435 filter, covering
$\lambda\lambda$5330--7540, at a dispersion of 0.7\,\AA/pixel, corresponding
to a resolution of R\,$\sim$\,1000 at 6300\,\AA. Two  
1800 sec exposures were taken before seeing conditions deteriorated. 
Identical flux standards were again used with this configuration.
For all datasets a standard data reduction was carried out using 
FIGARO\footnote{\tt 
www.starlink.rl.ac.uk/star/docs/sun86.htx/sun86.html}
i.e., bias subtraction, flat field correction, extraction, 
wavelength and flux calibration.

\section{Wolf-Rayet candidates}

\subsection{New Catalogue}

Our frames were positioned about one arcminute East of the centre of
NGC~300 such that they covered the bright H{\,\sc ii} regions Deh 53
and 137 (Deharveng et al.~\cite{deh88}). Therefore, we have surveyed 
the central region of the galaxy, but do not include
the outer spiral arms beyond a Holmberg radius $\rho_0 \gapprox 0.4-0.5$,
where $\rho_0$ = 9.75\,\arcmin$ \sim$ 5.75\,kpc. The 
sensitivity by which WR emission features can be detected obviously
depends on the signal/noise ratio. In this case the signal is the flux 
difference between on-off frames. We first identified the $\lambda$4684 
emission 
objects on the difference image. The selection criteria were a stellar
appearance and a peak intensity of at least 6$\sigma$. The probability
for any of the listed objects of having indeed a WR excess is therefore
very high. For these objects we subsequently picked the V, emission 
($\lambda$4684) and
continuum ($\lambda$4781) magnitudes from the (rather long) DAOPHOT 
photometric list. 

It should be noted that photometry is hampered by a variable 
background due to unresolved galaxy emission and heavy crowding, which
can be particularly severe in OB associations. It
follows that while the identification as a $\lambda$4684 emission object is
rather reliable, the quantitative measurement of a WR excess is less
certain. Our narrow-band images are complete down to 23.7 mag 
while the 
%distance modulus is 26.53 mag to NGC~300 (Freedman et al. 2001).  the 
3$\sigma$ detection limit was at 24.7 mag.

We present a catalogue of the 58 WR stars/candidates identified
in our images in Table~\ref{catalog}. We include spectral types taken
from the literature, updated in case of revisions from our new spectroscopy
(see Sect.~\ref{spectroscopy}), plus associated H\,{\sc ii} regions from
Deharveng et al. (1988).  De-projected galactocentric distances were
calculated using parameters from Table~1 of Deharveng et al. (1988).
In Figs.~\ref{nucnw.fig} to \ref{hiififi.fig} we give finding charts for 
the WR candidates. In all of them a horizontal bar is plotted that
represents 10\,\arcsec. The orientation is as usual: North to the top
and East to the left. All finding charts are from our 
$\lambda$4684 narrow-band filter images.

\subsection{Nature of candidates}

\begin{figure}
\epsfxsize=8.0cm \epsfbox[30 50 483 420]{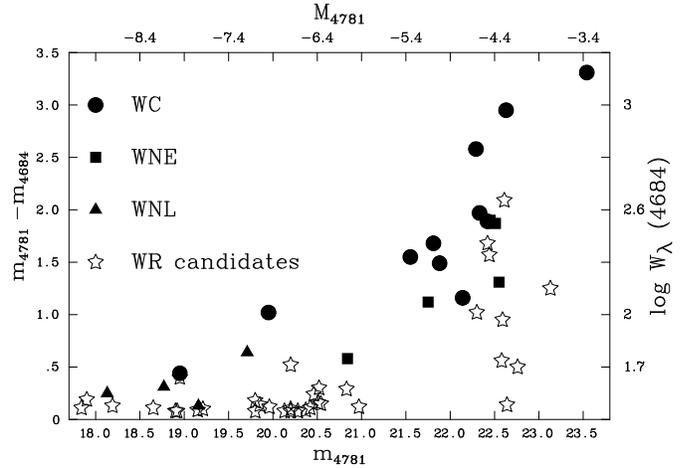}
\caption{Comparison between the $\lambda4781$ continuum magnitudes of
the WR candidates in NGC~300 and the $\lambda$4684 excess
(stars). Spectroscopically confirmed WR stars are presented
in the key. {\it Approximate} absolute magnitudes and line equivalent
widths are indicated, as
discussed in the text. Three WR stars are not shown in this figure --
\#1 (located at the edge of the image surveyed 
-- Fig.~\ref{nucnw.fig}), \#37 and \#51 (both are severely blended).}
\label{excess}
\end{figure}

39 out of our 58 WR candidates with a $\lambda$4684 excess are newly
identified in this study. We compare the $\lambda$4781 continuum magnitudes 
with the $\lambda$4684 excess in Fig.~\ref{excess}. 
Stars with previous (or new) spectroscopic confirmation are indicated, 
as shown in the key. We include estimates of the absolute
magnitudes at $\lambda$4781, assuming a distance modulus of 26.53 mag
to NGC~300 (Freedman et al. 2001) and $A_{\lambda4781}=$0.36 mag, 
corresponding to $E_{\rm B-V}$=0.10\,mag which is the mean interstellar
reddening towards H\,{\sc ii} regions of NGC~300 within $\rho/\rho_0$=0.5
derived by  Deharveng et al. (1988). 

Fig.~\ref{excess} clearly separates the large $\lambda$4684 
excesses of the visually faint WC and early-type WN stars,
from the small excesses of the visually bright 
late-type WN stars and WR binaries. {\it Approximate} line equivalent
widths (in \AA) are also presented in the figure, 
as estimated from stars in which optical spectroscopy is available
(see also Fig.4 of Massey \& Johnson 1998). Such comparisons would
represent the sole means by which WR populations might be identified
in galaxies which are too distant (or reddened) for confirmatory 
spectroscopy to be obtained, even with large 8--10m telescopes.

Previously known WR stars tend to have large $\lambda$4684 excesses 
of $\gapprox$ 1 mag, corresponding to emission equivalent widths of
$\gapprox$100\,\AA. Exceptions include those WR stars which lie in
well surveyed OB associations. We find six more WR candidates
with such a large WR excess (\#3, \#5, \#22, \#30, \#31 and \#37). 
They are rather faint with $m_{4684}>$\,20.5 mag,  which is 
presumably why they escaped earlier detection. From this sample, \#22 and 
\#30 were observed spectroscopically, such that both were 
confirmed as WR stars -- see Sect~\ref{spectroscopy}. 

Single, early-type WC stars have $\lambda$4650--4686
emission equivalent widths of $\gapprox$1000\,\AA, and are visually rather
faint. Consequently only three WC stars, with an excess of $\geq$2.5 mag
 are likely to be single, namely \#29, \#47 and \#48.
Two of these stars are discussed in detail in Sect.~\ref{analysis}. 
Other WC stars are almost certainly multiple, or suffer contamination
from stars along the same line-of-sight. Early-type WN stars possess
He\,{\sc ii} $\lambda$4684 emission equivalent widths of 
100--400\,\AA, corresponding to excesses of 1--2 mag. From Fig.~\ref{excess},
it is likely that \#28 and \#55 are single, whilst others are probably
multiple. For those remaining WR candidates with excesses greater than
1 mag, \#5 is a strong WC+O candidate, whilst the remainder are 
probable WN+OB systems.

There are also a handful of new WR candidates with
an excess in the range 0.3 to 1 mag (\#9, \#10, \#41, \#42, 
\#51, \#52 and \#54), corresponding to emission equivalent widths
in the range $\sim$30--100\,\AA. Their relatively small line strengths
suggests that if they are visually bright, with $M_{4781} <-6$, they are 
WR binaries (e.g. \#11) or single WN7--9 stars (e.g. \#38). If they are
visually faint, they are probably weak-lined single WN stars (e.g. \#41).
From this group, \#9 was observed spectroscopically
and confirmed as a WN star -- see Sect.~\ref{spectroscopy}.
Numbers 51 and 54 lie in the giant H\,{\sc ii}
region Deh\,137 while \#10 is another WR object in the nuclear
area of NGC\,300 (Fig.~\ref{nucnw.fig}). 

25 WR candidates have a $\lambda$4684 excess below 0.3 mag. 
It is possible that some of these stars are borderline WR/Of stars. 
Except in one case, they are visually bright, with $M_{4781} <-$6,
as expected for a WR binary, single late WN
star, or an extreme O-type supergiant, with strong 
He\,{\sc ii} $\lambda$4686 emission (i.e. an Of star). Indeed, 
\#6 was independently found by Bresolin et al. (2002a) and 
classified as a WN\,11 star. Our $\lambda$4684 excess of 0.13 mag is 
in close agreement with the spectroscopy of Bresolin et al. (2002b). 
This indicates that a He\,{\sc ii} excess of $\sim$\,0.1 mag can be 
reliably measured with this instrumentation, and hence also that 
Of stars and very late WN stars can easily be detected. Most candidates
from the present sample are probably WR rather than Of since, since 
by definition,
He\,{\sc ii} $\lambda$4686 equivalent widths of the latter do not
exceed $\sim$12\,\AA\ (Bohannan \& Crowther 1999).

\subsection{Completeness, Surface Density and the WC/WN Ratio}

In addition to verifying the likely-hood of whether our candidates
are genuine WR stars, how  complete is our survey?
The continuum filter centred at 4781\,\AA\ lies midway 
between the usual Smith (1968)
WR narrow-band $b$ (4270\,\AA) and $v$ (5160\,\AA) filters.
Typical intrinsic colours of WC and WNE stars are 
$(b-v)_0\sim-0.2\pm0.1$, such that
one would expect $b-v\sim-0.1\pm0.1$ mag for WR stars in NGC~300, given
typical extinctions of $E(b-v)\sim0.1$ mag (equivalent to $E(B-V)=0.12$\,mag).
Consequently, continuum filter
measurements should correspond closely (within $\sim$0.1 mag) to $b$ or
$v$ magnitudes. i.e. a completeness to $v$=23.7 mag 
will be equivalent to $M_{v}\sim-3.2$ mag.
According to Table~28 from van der Hucht (2001), 94\,\% of the 227 known
Galactic WR stars are brighter than
$M_{v}$=$-$3.5 mag, so our census of the central region of NGC~300
should be reasonably 
complete, except those suffering from high visual extinction.
\begin{figure}
\mbox{\psfig{figure=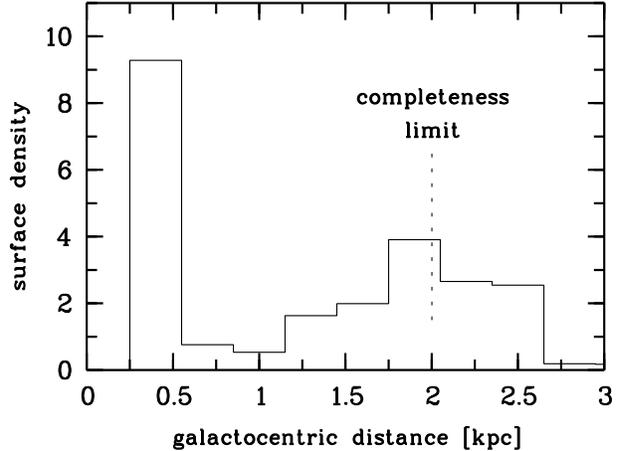,width=8.5cm,height=6.5cm,clip=}}

\caption[]{Surface density of WR stars/kpc$^2$ in the inner region of
NGC~300. The completeness limit is reached at 2kpc ($\rho/\rho_O$=0.4), 
since our imaging was not centred on the nucleus.}
\label{surfdens}
\end{figure}

\begin{figure}
\epsfxsize=7.0cm \epsfbox[50 215 403 600]{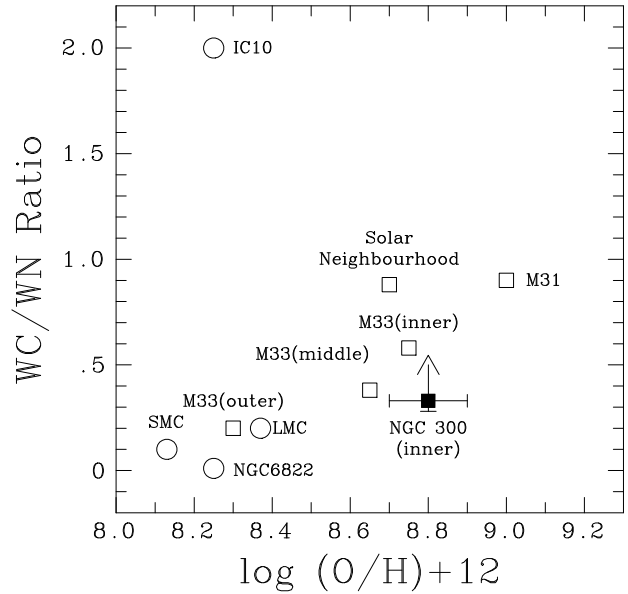}
\caption{The relative number of WC and WN stars in Local Group
galaxies versus metallicity (Massey 2003), supplemented by NGC~300
(solid) from the present work. The WC/WN ratio for IC10
is probably an overestimate.}
\label{WC_WN}
\end{figure}

\begin{figure*}
  \hfill
  \begin{minipage}[t]{.45\textwidth}
    \begin{center}  
      \epsfxsize=8cm \epsfbox[100 50 543 740]{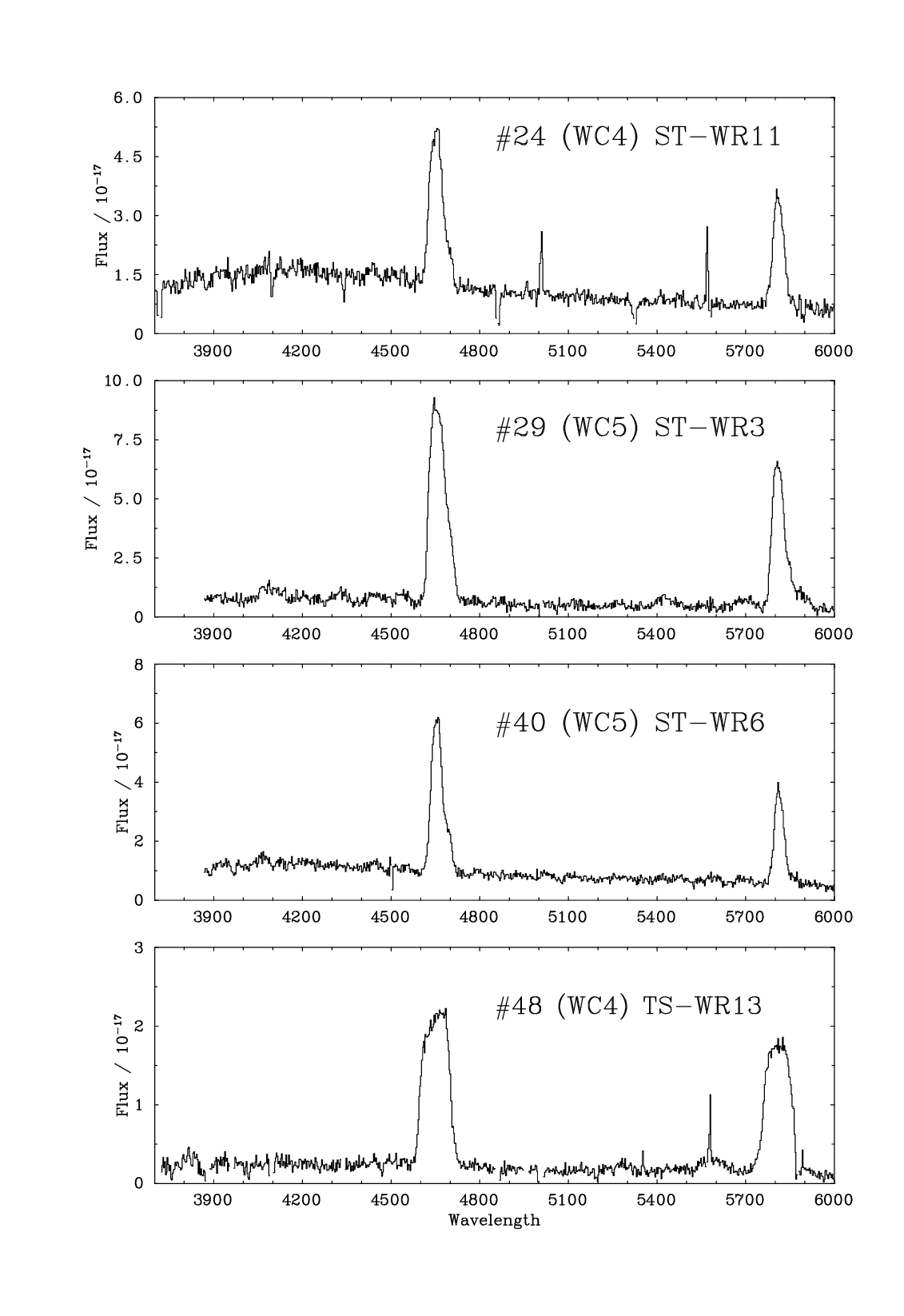}
      \caption{Low dispersion FORS2 spectroscopy of four previously
               identified WR stars in NGC~300.}
      \label{oldwr}
    \end{center}
  \end{minipage}
  \hfill
  \begin{minipage}[t]{.45\textwidth}
    \begin{center}  
      \epsfxsize=8cm \epsfbox[100 50 543 740]{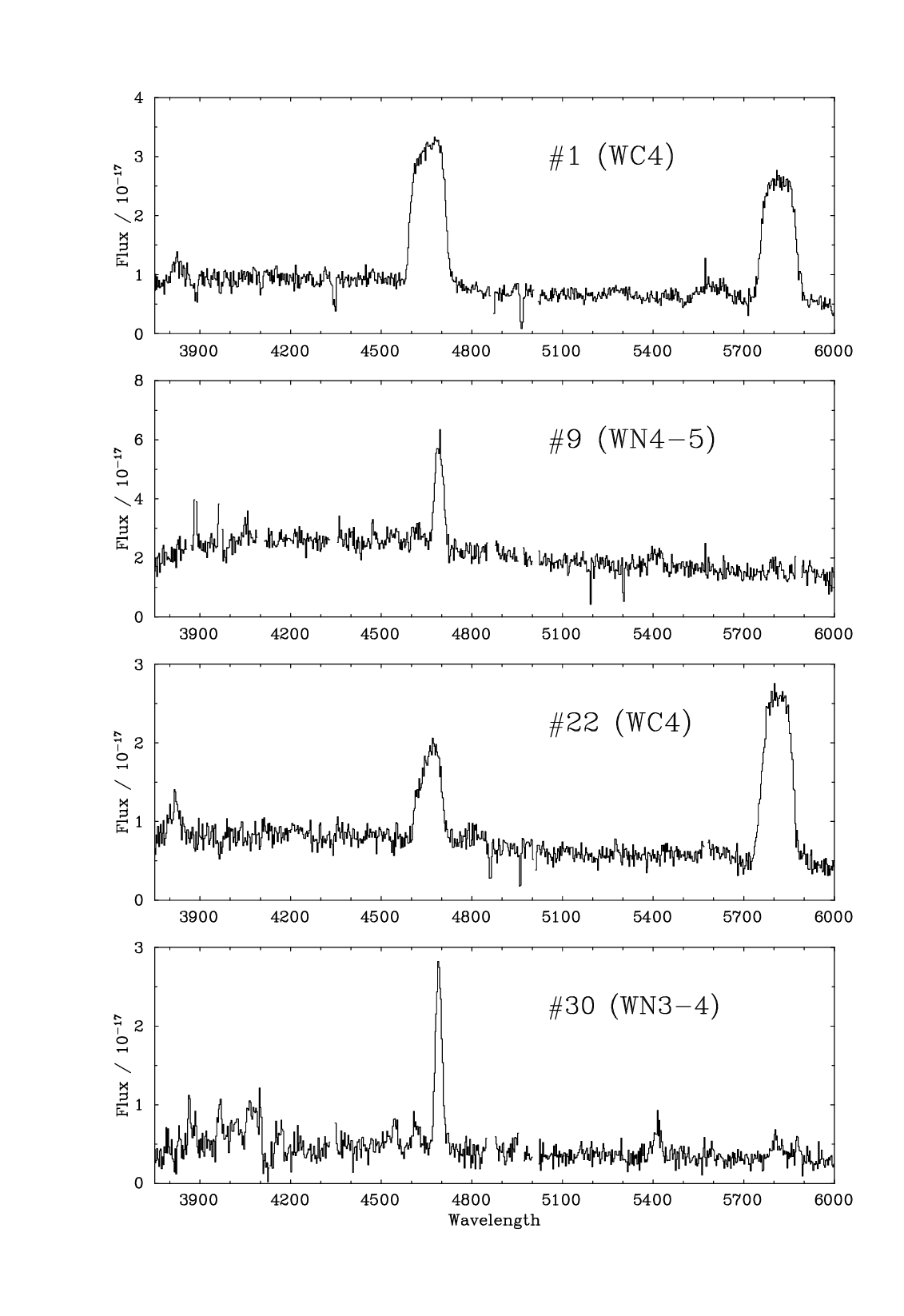}
      \caption{Low dispersion FORS2 spectroscopy of four newly
               identified WR stars in NGC~300.}
      \label{newwr}
    \end{center}
  \end{minipage}
  \hfill
\end{figure*}

In Fig.~\ref{surfdens} we plot the WR surface density versus
the galactocentric distance. The WR distribution in the nuclear region
is particularly interesting. While the very centre is apparently free of 
WR stars we find a sharp increase of the surface density at a 
galactocentric
distance of about 0.4\,kpc. Further outside it drops first to a minimum at 
around 1\,kpc and rises again outwards. Qualitatively, a similar
behaviour is observed in our galaxy (van der Hucht 2001) but in NGC~300
the drop is much shallower, about $-$0.3 dex between 0.5 and 2\,kpc 
instead of $-$1.5 in the galaxy. The highest surface density in NGC~300 
occurs in the
Deh 137 H{\sc ii} region, alias OB association AS 102 which contains
15 WR stars in an area that spans 0.3$\times$0.3\,kpc implying a WR
density of about 150 WR stars/kpc$^2$. Massey \& Johnson (1998) compare
WR surface densities of other Local Group galaxies, such that WR surface 
densities range from 1/kpc$^2$ in the SMC, to 2 in the LMC and $\sim$4 
in M33. 

As discussed in the introduction, the WC/WN ratio for NGC~300 
prior to the present study, i.e. $\sim$2, 
was unusually high relative to more complete
surveys of Local Group galaxies. Ideally, one might use additional
narrow band filters at (C\,{\sc iv}) $\lambda\lambda$5801-12 plus 
a nearby continuum region to discriminate between WN and WC stars, as
recently carried out by Royer et al. (2001) for IC10. In the absence of
such filters, we have been able to infer likely WN or WC subtypes
for those stars without spectroscopy from inspection of Fig.~\ref{excess}. 
We suggest that at least 13 WR stars in NGC~300 are WC stars, 
i.e. 12 for which spectroscopy is available, plus \#5. A further 3 {\it may} 
host WC binaries, namely \#12, \#31, \#33, such that the WC/WN ratio
for the central regions of NGC~300 is $\geq$12/46=0.3, or more likely
$\sim$15/43=0.35. This falls close
to that observed in comparable regions of M33, according to Massey (2003),
as illustrated in Fig.~\ref{WC_WN}.

\section{Spectroscopy of Wolf-Rayet stars}\label{spectroscopy}

\subsection{Previously identified WR stars}

We present flux calibrated low dispersion FORS2 spectroscopy of four 
previously observed NGC~300 WC stars in Fig.~\ref{oldwr}. 
These datasets are 
superior to previous 4m observations, and so allow us to obtain revised
spectral types, using the scheme of Crowther et al. (1998).

We revise the original classification for \#29, alias WR3 
(Schild \& Testor 1991), from WC4--5 to WC5 since our dataset reveals
weak C\,{\sc iii} $\lambda$5696, with 
$W_{\lambda}$($\lambda$5696)/$W_{\lambda}$($\lambda\lambda$5801-12)$\sim$0.1
and O\,{\sc iii-v} $\lambda$5592 weak/absent. This star is probably single,
given that its emission line spectrum is comparable in strength (e.g.
$W_{\lambda}$($\lambda\lambda$5801-12)$\sim$800\,\AA) to 
apparently single Galactic and LMC WCE stars.

A firm classification is possible
for \#40, alias WR6 (Schild \& Testor 1992) for which we also assign WC5
(updated from WC4--6) since
C\,{\sc iii} $\lambda$5696 is again present, with
a similar strength to O\,{\sc iii-v} $\lambda$5592 and 
$W_{\lambda}$($\lambda$5696)/$W_{\lambda}$($\lambda\lambda$5801-12)$\sim$0.1.
\#40 is almost certainly multiple, since 
$W_{\lambda}$($\lambda$5801-12)$\sim$230\,\AA.

Testor \& Schild (1993) previously assigned a WC5 spectral type for
\#24 (their WR11), which we revise to WC4, given that
$W_{\lambda}$($\lambda$5696)/$W_{\lambda}$($\lambda\lambda$5801-12)$\leq$0.05.
C\,{\sc iv} $\lambda$5801--12 is again unusually weak, with
$W_{\lambda}$($\lambda$5801-12)$\sim$200\,\AA\ indicating either
binarity or a line-of-sight companion.

The spectral appearance of \#48, alias WR13 (Testor \& Schild 1993), alias
IV-3 (Breysacher et al. 1997) is in marked contrast to the other WCE stars
whose spectroscopy is presented here, with much broader lines -- 
FWHM($\lambda\lambda$5801-12)$\sim$86\,\AA\
versus 36--47\,\AA. Breysacher et al. (1997) interpreted this large FWHM
as an indication of a (rare) WO subtype, which possess strong 
O\,{\sc vi} $\lambda\lambda$3811-34 emission lines, and assigned a WO4 spectral type
, whilst our spectroscopy reveals that O\,{\sc vi} is weak/absent in \#48.
Since C\,{\sc iii} $\lambda$5696 is also absent, a WC4 spectral type 
is appropriate. Willis et al. (1992) discuss problems with using
FWHM as indicators of spectral type for WC stars in M33. We suspect that
\#48 is single since $W_{\lambda}$($\lambda\lambda$5801-12)$\sim$1500\,\AA.

\subsection{Newly identified WR stars}

We present optical spectroscopy of four newly identified NGC~300
WR stars in Fig.~\ref{newwr}, two WN and two WC stars. The WC stars
\#1 and \#22 are rather similar. They have lines widths which are higher
than \#48, with FWHM($\lambda\lambda$5801--12)$\sim$91-100\,\AA, and similar
line strengths, 
$W_{\lambda}$($\lambda\lambda$5801-12)$\sim$500\,\AA. WC4 subtypes
are appropriate for both stars since
there is no evidence of C\,{\sc iii} $\lambda$5696, with O\,{\sc vi} 
$\lambda\lambda$3811-34 weak. Both stars are probably binaries.

The two WN stars \#9 and \#30 are early-type, 
since N\,{\sc v} $\lambda\lambda$4603-20
is prominent, with N\,{\sc iii} $\lambda\lambda$4634--41 weak (\#9) or absent (\#30).
Following the classification scheme of Smith et al. (1998) one obtains
a spectral type of WN4--5 for \#9 (N\,{\sc iv} $\lambda$4058 $\sim$
N\,{\sc v} $\lambda\lambda$4603-20), and WN3--4 for \#30 (the region around
N\,{\sc iv} $\lambda$4058 is noisy). 
One cannot use the (primary) He\,{\sc i-ii} classification
diagnostics for these stars due to the strong nebular contamination, and
weak He\,{\sc i} $\lambda$5876 emission.

\begin{figure}
\epsfxsize=7.0cm \epsfbox[20 150 463 640]{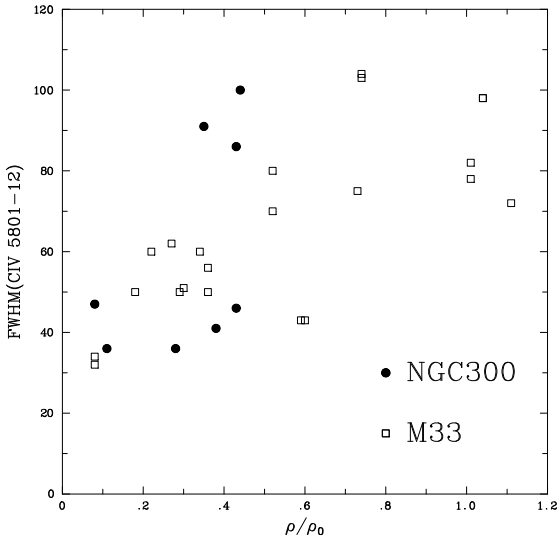}
\caption{Comparison between FWHM(C\,{\sc iv} $\lambda\lambda$5801-12), in \AA,
and galactocentric distance for early WC stars in NGC~300 (solid) and
M~33 (open, Willis et al. 1992), as a fraction of the Holmberg radius.}
\label{fwhm}
\end{figure}

\subsection{WC line width}

Willis et al. (1992) identified a correlation between line width
(FWHM C\,{\sc iv} $\lambda\lambda$5801-12) and galactocentric
distance for WCE stars in M33 in the sense that stars at larger
galactocentric distance (i.e. lower metallicity) had broader lines
than those in the nucleus (with higher metallicity). We present
our measurements for 6 WC stars in NGC~300 in Fig.~\ref{fwhm}, supplemented
by data from Schild \& Testor (1991, 1992) for \#14 (WR1) and \#47 (WR5),
and including data from Willis et al. (1992) and references therein for M33.
For NGC~300, there is a very large scatter
in FWHM at $\rho/\rho_0\sim0.4$, arguing against a tight correlation in
general, although the present results are in favour of a deficit of broad-lined
WC stars in the nucleus.
Nevertheless, 
firm conclusions await spectroscopy of larger numbers of WC stars in both
galaxies.

\section{Analysis of WC stars}\label{analysis}

As discussed above in Sect.~\ref{spectroscopy}, 
two WC stars in NGC~300 are apparently single, and
have sufficient quality observations for detailed analyses to be carried
out. Ultimately, large numbers of WR stars need to be studied in 
galaxies spanning a wide range metallicities to place adequate constraints
on evolutionary models. Recent studies, using identical techniques, 
have been presented for single WC stars in the Milky Way 
(e.g. Dessart et al. 2000), LMC (Crowther et
al. 2002), M31 (Smartt et al. 2001) and M33 (Abbott et al. 2003). We
now proceed to study \#29 (WC5), located close to the nucleus of NGC~300 
with a probable metallicity of $\sim Z_{\odot}$ 
according to  $10^4 {\rm O/H} = 7.5 - 5.3 \rho/\rho_0$
(Deharveng et al. 1988),  and \#48 (WC4),
located at $\rho/\rho_0$=0.43 with $\sim\,0.6\,Z_{\odot}$.

\subsection{Technique}

We employ the non-LTE code of Hillier \& Miller (1998), which 
iteratively solves the transfer equation in the co-moving frame 
subject to statistical and radiative equilibria in an expanding, spherically 
symmetric and steady-state atmosphere. 
Specific details of the (extremely complex) He, C, O, Ne, Si, P, S, Ar, Fe 
model atoms used for our quantitative analysis are 
provided in Crowther et al. (2002).  
We assume that the wind is clumped with a volume filling factor, $f\sim0.1$.
 We parameterise the filling factor 
so that it approaches unity at small velocities.

As usual, a series of models were calculated in which stellar 
parameters ($T_{\ast}$, log\,$L/L_{\odot}$, $v_{\infty}$
$\dot{M}/\sqrt{f}$, C/He, O/He) were adjusted until 
the observed ionization balance, line strengths, widths, 
and absolute $v-$ band flux were reproduced. 
Because of the substantial effect 
that differing mass-loss rates,  temperatures and elemental abundances 
have  on the emergent spectrum, this was an iterative process.
Abundances other than He, C and O, namely 
Ne $\ldots$ Fe, were set at 0.5$Z_{\odot}$, although the exact
choice does not greatly affect the emergent spectrum (Crowther et al. 2002).
As discussed above, we select $v\sim m_{4781}$ for both stars. Despite
the narrow spectral range available, we derive
a reddening of $E_{b-v}$=0.08\,mag
($E_{B-V}$=0.10\,mag) for \#29 and $E_{\rm b-v}$=0.12\,mag ($E_{B-V}$=0.15\,mag)
for \#48, in good agreement with H\,{\sc ii} regions (Deharveng et al. 1988).
A standard Galactic extinction law of Seaton (1979) is adopted. 
Accounting for uncertainties in $v$, distance and reddening, we 
obtain 
$M_{v}=-$4.6$\pm$0.2\,mag for \#29 and $M_{v}=-$3.5$\pm$0.2\,mag for \#48.

The wind ionization balance is ideally selected on the basis of isolated
optical lines from adjacent ionization stages of carbon 
and/or helium, e.g. He\,{\sc i} $\lambda$5876/He\,{\sc ii} $\lambda$5412.
In practice, this was 
difficult to achieve because of the severe blending, so our
derived temperature should be treated as approximate. Detection
of He\,{\sc i} $\lambda$5876 appears to be robust in \#29, due to
its relatively low wind velocity, whilst there is an ambiguity in
this feature for \#48, since it is possible that the observed feature
represents the electron scattering wing of C\,{\sc iv} for which
we have adopted the filling factor, $f$.  We also simultaneously match
C\,{\sc iii} $\lambda$6740 and C\,{\sc iv} $\lambda$5801, the former selected
in preference to C\,{\sc iii} $\lambda$5696 which is very sensitive
to the exact 
ionization structure (Hillier \& Miller 1998; Crowther et al. 2002).
The standard C/He diagnostic,  
He\,{\sc ii} $\lambda$5412/C\,{\sc iv} $\lambda$5471, was used  
since their relative strengths are insensitive to temperature and mass-loss.
Oxygen abundances were difficult to constrain, since we relied solely
on O\,{\sc iii-v} $\lambda$5592 (Crowther et al. 2002). 
Consequently, caution is advised  when comparing the present 
O/C determinations with (Galactic and LMC) WC stars for 
which the superior $\lambda\lambda$2800--3100 diagnostics are
available.

\begin{figure}
\epsfxsize=8.0cm \epsfbox[65 125 465 725]{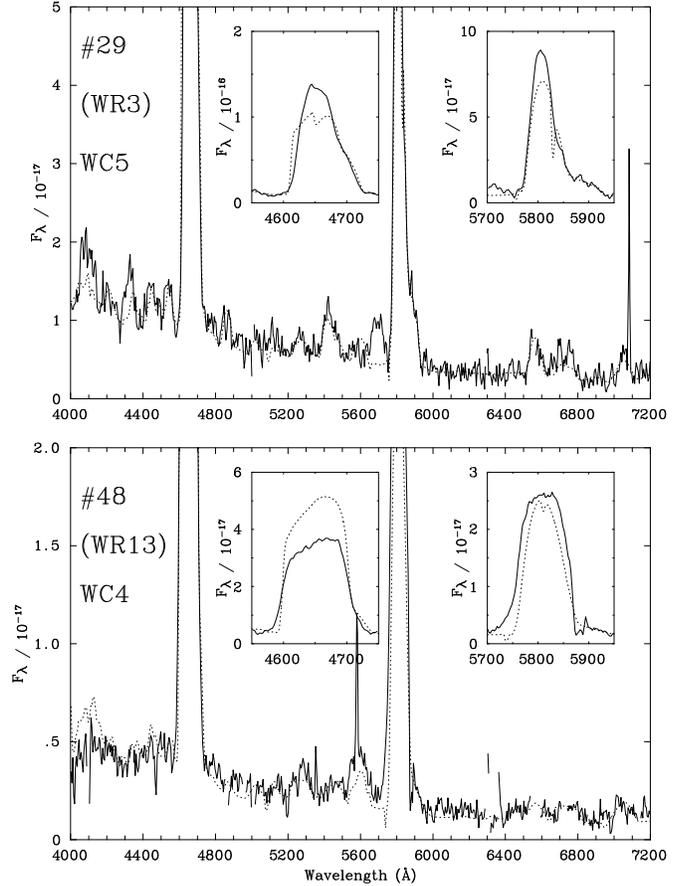}
\caption{{\it Upper panel:} Synthetic spectral fit (dotted)
to FORS2 observations (solid) of NGC~300 \#29 (WR3,
WC5), de-reddened by $E_{B-V}$=0.10\,mag. Close up views of the
C\,{\sc iii-iv}-He\,{\sc ii} $\lambda\lambda$4650--4686 and 
C\,{\sc iv} $\lambda\lambda$5801-12 regions are indicated.
{\it Lower panel:} Same for NGC~300 \#48 (WR13, WC4) for
a reddening of $E_{B-V}$=0.15\,mag.}
\label{ngc300-wc}
\end{figure}

\subsection{Results for NGC300 \#29 (WR3, WC5)}

Our FORS2 spectroscopic data of \#29 is shown in the upper 
panel of Fig.~\ref{ngc300-wc}.
Overall, the spectrum is reasonably well reproduced by our model fit,
except that the C\,{\sc iii}  $\lambda$5696 profile is strongly underestimated,
whilst C\,{\sc iv} $\lambda\lambda$5801--12 is 40\% too weak.
From our recent experience it is difficult to simultaneously reproduce
the strength of C\,{\sc iii} $\lambda$5696 feature together with other
diagnostics in early WC stars. We find $T_{\ast}\sim$100\,kK, 
log$(L/L_{\odot})$=5.5,  $v_{\infty}\sim 2700$\,km\,s$^{-1}$, and 
$\dot{M} \sim10^{-4.6}\,M_{\odot} {\rm yr}^{-1}$.
We estimate C/He$\sim$0.08 by number from He\,{\sc ii} 
$\lambda$5412/C\,{\sc iv} $\lambda$5471. 
The weak O\,{\sc iii-v} $\lambda$5592 feature suggests
a low oxygen content of O/He$\le$0.05 by number.

\subsection{Results for NGC~300 \#48 (WR13, WC4)}

We compare our spectroscopy of \#48 with our synthetic
model in the lower panel of Fig.~\ref{ngc300-wc}. Again, reasonably
good agreement  is achieved, although the broad emission lines of \#48
hinder detailed comparisons. The blend comprising principally C\,{\sc iii} 
$\lambda$4647--51, C\,{\sc iv} $\lambda$4660 and He\,{\sc ii} 
$\lambda$4686 is rather too strong in the synthetic model. Our derived
parameters are  $T_{\ast}\,\sim$\,95\,kK, log$(L/L_{\odot})$=5.2, 
$v_{\infty}\,\sim\,3750$\,km\,s$^{-1}$, and  
$\dot{M}\,\sim\,10^{-4.8}\,M_{\odot} {\rm yr}^{-1}$.
We estimate C/He\,$\sim$\,0.5 by number, although this
ratio should be treated with caution, given the poor
quality of the observations -- recall \#48 has the
faintest continuum ($v$=23.5 mag) 
of all 58 WR candidates in NGC~300.
The O\,{\sc iii-v} $\lambda$5592 feature suggests
a high oxygen content of O/He$\ge$0.1 by number.

\subsection{Comparison with WC stars in the Galaxy and LMC}

\begin{figure}
\vspace{11.0cm}
\includegraphics{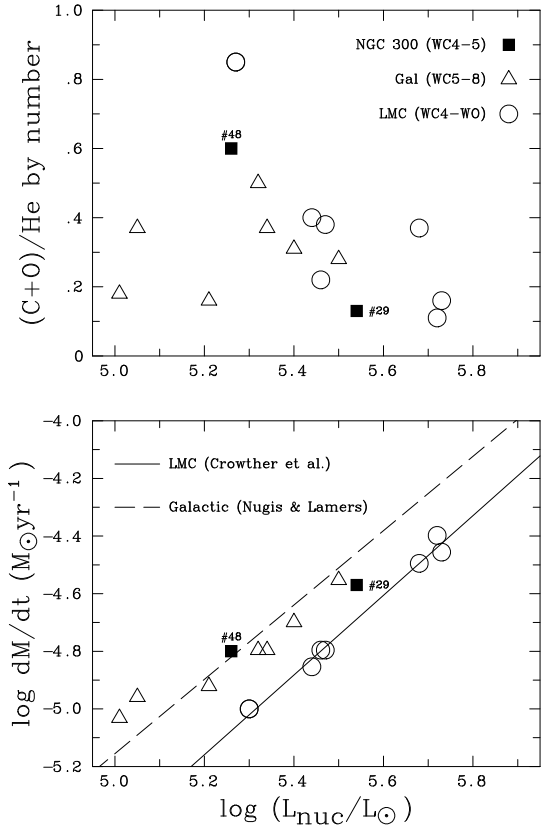}
\caption{{\it Upper panel:} Comparison between the (C+O)/He ratios of WC stars 
in NGC~300 (filled symbols, this work), the Milky Way and LMC 
(see Crowther et al. 2002), as indicated in  the key; {\it Lower panel:}
Comparison between the mass-loss rates and luminosities
of NGC~300 (filled symbols, this work), Milky Way and LMC WC stars
(see Crowther et al. 2002), as indicated above. A fit to the
LMC WC4 stars (solid) is indicated, as is the generic Galactic
WR mass-loss calibration from Nugis \& Lamers (2000), 
assuming C/He=0.2, C/O=4 by number.}
\label{c+o_mdot}
\end{figure}

Crowther et al. (2002) recently contrasted the properties of Solar
neighbourhood 
and LMC WC stars, to which we can now add NGC~300 \#29 and \#48.
The upper panel in Fig.~\ref{c+o_mdot} compares (nuclear) luminosities and 
(C+O)/He abundances
for WC stars in the three galaxies. Nuclear luminosities are derived
by taking into account the wind blanketing effects discussed by
Heger \& Langer (1996). In contrast with the results of Heger \& Langer,
who indicated revisions of up to 0.3 dex in luminosity, 
revised mass-loss rates due to clumping
yield rather small corrections,  typically 0.05 dex.
Current  masses of 16.3 and 11.6\,$M_{\odot}$ are
determined for \#29 and \#48, respectively.  Crowther et al. (2002) found
that low metallicity (LMC) WC stars possess higher luminosities
than those at high metallicity (Milky Way). 
This can be explained since one would require a higher initial mass cut-off, 
for a massive star to progress through to the WC stage at low
metallicity, because of reduced mass-loss rates during the main-sequence
and post-main sequence evolution. The small sample of NGC~300 stars does
not allow firm conclusions to be drawn, given that they span a 
range in luminosity common to LMC and Galactic WC stars.
% langer_mdot.exe

Fig.~\ref{c+o_mdot} suggests that \#48 has a 
high (albeit uncertain) metal content, whilst
\#29 has a (C+O)/He ratio which is 
amongst the lowest of WC stars studied to date. 
Analyses of larger numbers of WC stars in external 
galaxies is urgently required for general trends to be identified.

Crowther et al. (2002) also identified a trend towards weaker winds
for WC stars at lower metallicities, such that $\dot{M} \propto Z^{0.5}$,
which explained the relative WC subtype distributions in the LMC and
Galaxy. The lower panel of Figure~\ref{c+o_mdot} compares luminosities and 
mass-loss rates of Galactic, LMC and NGC~300 WC stars, including a fit to the LMC 
data (from Crowther et al.) plus the Galactic calibration of Nugis \& Lamers
(2000). Both \#29 (WC5) and \#48 (WC4) lie on, or slightly below,
the Galactic calibration, as would be expected from the metallicities 
inferred from their  locations in NGC\,300 (Deharveng et al. 1988; Zaritsky et 1994). 
Of course, the difference between Galactic and LMC calibrations amounts to less than a 
factor of two  in mass-loss rate, and  Smartt et al. (2001) discuss, in detail, 
limitations with adopting generic H\,{\sc ii} region abundances 
with regard to WC properties in M31.

\section{Conclusions} 

We have demonstrated the feasibility of using narrow-band filters to
detect Wolf-Rayet candidates in NGC~300. 
Restricting our survey to
the central 6.8\,\arcmin$\times$6.8\,\arcmin\ region, we have trebled 
the known 
WR content from 20 to 58 stars, within a factor of two of the
global content of its northern Local Group counterpart, M33. 
Surveys of the outer spiral arms of NGC~300 are sought in order to determine its
total WR content, which probably approaches $\sim$\,100, 
as is spectroscopic confirmation of remaining candidates.
The WC/WN ratio of the central region of NGC~300 has been
revised from $\sim$\,2 to $\sim$\,1/3, in reasonable accord with Local 
Group galaxies spanning a similar metallicity range. 
Modern  abundance analyses of H\,{\sc ii} regions and/or AB supergiants
are urgently required to verify previous determinations of the 
metallicity gradient of NGC~300. We have purposefully not discussed the
WR/O ratio in NGC~300, since it is extremely difficult to 
constrain  this ratio observationally,  as discussed by Massey (2003).

Using VLT-FORS2, 600 sec imaging provides over 90\,\% completeness for WR
candidates down to an excess of 0.1\,mag at a distance of 2\,Mpc. The number
of known WR stars can therefore be rapidly increased with a moderate
investment of observing time.  This approach greatly improves our chance 
of witnessing a WR star undergoing a  supernova explosion in the nearby 
universe. Imaging surveys towards such goals are presently 
underway (e.g. Smartt et al.  2002), albeit based solely around broad-band 
filters, such that WR candidates can not easily be identified. Under 
favourable conditions one can reasonably expect to extend our 
imaging/spectroscopic  approach to WR stars within galaxies  at distances 
of up to at least $\sim$\,5\,Mpc. Recession radial velocities, 
shifting WR emission lines redward of the $\lambda$4684 filter, 
only become problematic in excess of 1000 km\,s$^{-1}$,  typically 
corresponding to distances in excess of $\geq$10\,Mpc.
We have recently obtained VLT-FORS2 narrow-band imaging of
M83 (NGC~5236), at distance of 3.2\,Mpc\footnote{We assume that its 
distance is comparable to NGC~5253} (Freedman et al. 2001).
M83 has a metallicity of up to  5\,$Z_{\odot}$ (Webster \& Smith 1983), 
such that a large WR population is expected. Indeed, WR signatures are  
already known from H\,{\sc ii}  region studies (Bresolin \& Kennicutt 
2002).

Once the census of WR stars is reasonably complete in a galaxy or part
thereof, we can obtain surface density plots. We present here the
WR star distribution in the central 2\,kpc of NGC~300. We find that
the very centre of the galaxy is apparently void of WR stars, in contrast
with our own Galaxy, but that 
a maximum of the surface density occurs at a galacto-centric distance of 
about 0.4\,kpc. At 1\,kpc the surface density drops to a minimum, beyond
which it steadily increases to about half the value of the
0.5\,kpc ring. NGC~300 compares favourably with most Local Group galaxies
in its WR surface density, with the exception of IC10, and 
perhaps M33 (Massey \& Johnson 1998).

We have also illustrated that single, early-type WC stars with 
$v\,\sim$\,23 mag can be quantitatively 
studied using modest integration times  with 
VLT-FORS2. More problematic is the challenge of obtaining 
uncontaminated WR spectroscopy at such large distances, since a slit width of 
1\,\arcsec\ corresponds to a spatial scale of $\sim$\,10\,pc at 2\,Mpc. 
% versus 0.25pc at the distance of the LMC. 
Isolated WR stars are present, although they are
in the minority and will be even more problematic for still more distant 
galaxies. An order-of-magnitude reduction in slit size is ultimately
required using ground-based telescopes, 
without  a corresponding loss of throughput.

\begin{acknowledgements}
Thanks to John Hillier for the generous use of his model atmosphere
code and to the Conception group for providing photometric standards
in advance of publication. PAC 
acknowledges financial support from the Royal Society.
The routine to determine de-projected galactocentric distances was
adapted from fortran code provided by Phil Massey. 
\end{acknowledgements}

\end{document}